# Direction-Aware Masked Pretraining on 3D Seismic Data with Transfer to Cross-Area Acoustic Impedance Inversion

**Man Tang[1], Zhaoyun Zong[1], Diqiong Jiang[2], Gen Li[1] and Jiayun Li[1]**

1 State Key Laboratory of Deep Oil and Gas, China University of Petroleum (East China), Qingdao 266580, P. R. China

2 College of Computer Science and Technology, China University of Petroleum (East China), Qingdao 266580, P. R. China

## Abstract

Seismic feature extraction and acoustic impedance inversion are important for subsurface characterisation, but sparse well-log data limit the generalisation of data-driven inversion models across different areas. Self-supervised masked pretraining offers a way to exploit large volumes of unlabelled seismic data; however, conventional approaches often overlook the directional characteristics of 3D seismic data. We propose a direction-aware masked autoencoder (DA-MAE) that distinguishes lateral reflector structure from vertical waveform characteristics in 3D post-stack data. The framework incorporates this distinction into token representation, masking geometry, and reconstruction constraints on reflector continuity and waveform fidelity. We evaluate DA-MAE through masked reconstruction on an independent field survey and cross-area acoustic impedance inversion using two additional field areas. Reconstruction experiments show that performance is more sensitive to lateral token resolution than to moderate changes in vertical patch length, and that increasing encoder capacity does not fully compensate for coarse tokenization. Moreover, the preferred token scales and masking strategies vary between reconstruction and inversion, suggesting that reconstruction fidelity alone is not a reliable indicator of transferability. In cross-area inversion, the pretrained representations remain effective in the target area and improve impedance prediction, demonstrating their transferability across different seismic surveys. These results provide practical guidance for seismic-specific masked pretraining and its application to acoustic impedance inversion.

## 1. Introduction

Reliable extraction of seismic features and acoustic impedance inversion are important for quantitative subsurface characterisation (Yilmaz 2001; Chopra & Marfurt 2005). Post-stack impedance inversion uses seismic data to estimate subsurface impedance variations, while well-log information is commonly used to constrain the low-frequency background (Oldenburg *et al*. 1983; Latimer et al. 2000; Meng *et al*. 2024). Deep-learning methods have increasingly been introduced to learn nonlinear relationships between seismic responses and subsurface properties (Das *et al*. 2019; Khosro Anjom *et al*. 2024; Sun *et al*. 2024). In particular, semi-supervised and physics-guided strategies have been developed to make better use of sparse well information and geophysical constraints (Alfarraj & AlRegib 2019; Sun *et al*. 2021; Song *et al*. 2022; Sun & Zong 2024). However, supervised inversion still depends strongly on paired seismic and well-log data (Das *et al*. 2019; Li *et al*. 2025). In field surveys, impedance logs are sparsely distributed and provide only limited supervision for large 3D seismic volumes (Song *et al*. 2022; Zou *et al*. 2024; Li *et al*. 2025). This limits model training and can reduce generalisation across areas with different geological and acquisition conditions (Wu *et al*. 2020; Sheng *et al*. 2025).

Self-supervised pretraining provides a practical way to exploit large volumes of unlabelled seismic data before adaptation to inversion tasks. Masked autoencoders (MAEs) learn representations by reconstructing missing patches from partially observed inputs and have shown strong transferability in computer vision (He *et al.* 2022). This idea has subsequently been introduced into geophysical model building and seismic inversion. Li *et al.* (2023) used MAE-based Vision Transformer pretraining to improve subsurface model building with limited well information, while Dou and Li (2024) developed SeisMAE for 3D seismic acoustic impedance inversion. More recently, Li *et al.* (2025) combined MAE pretraining with fine-tuning for seismic–well subsurface model building, Sheng *et al.* (2025) demonstrated transferable representations using a large-scale seismic foundation model, and Wu *et al.* (2026) applied MAE-based seismic–well fusion to field 3D velocity model building. Together, these studies show that masked pretraining can learn transferable seismic representations while reducing reliance on densely labelled data.

Despite these advances, masked pretraining for 3D post-stack seismic data has rarely considered the different roles of the lateral and vertical directions. The lateral dimensions mainly characterise reflector geometry and spatial continuity, whereas the vertical direction contains waveform information along seismic traces. Treating these directions in the same way may therefore be suboptimal for seismic representation learning. In addition, random masking may leave sufficient neighbouring information for reconstruction, while pointwise losses mainly constrain local amplitude errors and provide limited control over reflector continuity and waveform characteristics. These issues suggest that tokenization, masking, and reconstruction should be designed specifically for the directional properties of seismic data.

To address these issues, we propose a direction-aware masked autoencoder (DA-MAE) for self-supervised representation learning from 3D post-stack seismic data. The framework incorporates the directional distinction between lateral reflector structure and vertical waveform characteristics into token representation, masking geometry, and reconstruction constraints. Anisotropic tokenization and direction-aware representation are used to preserve seismic structure at different scales, while trace-aligned tube masking encourages the model to exploit broader cross-trace context rather than local interpolation. A seismic-specific reconstruction objective further constrains reflector continuity and waveform fidelity. The learned representations are evaluated through masked reconstruction on an independent field survey and cross-area acoustic impedance inversion.

The main contributions of this study are summarized as follows:

(1) Direction-aware seismic representation. We develop a direction-aware masked pretraining framework for 3D post-stack seismic data that explicitly accounts for the different roles of lateral reflector structure and vertical waveform information.

(2) Controlled analysis of representation design. We systematically investigate the effects of token scale, masking strategy, and model capacity on seismic reconstruction. The results show that reconstruction is more sensitive to lateral token resolution than to moderate changes in vertical extent.

(3) Transfer beyond reconstruction. We evaluate the learned representations through cross-area acoustic impedance inversion and find that the token scales and masking strategies that perform well for reconstruction do not necessarily yield the best inversion results. This highlights the importance of evaluating seismic pretraining in the context of the downstream task.

## 2. Methods

### 2.1. Framework overview

DA-MAE consists of two stages: self-supervised pretraining and downstream transfer. During pretraining, the encoder learns representations from unlabelled 3D post-stack seismic volumes through masked reconstruction. For downstream inversion, the pretrained encoder is combined with a trace-oriented prediction head and jointly fine-tuned for acoustic impedance prediction. Figures 1 and 2 illustrate the pretraining and transfer stages, respectively.

Let the input seismic volume be denoted by $X \in \mathrm{R}^{C \times H \times W \times D}$ , where $C$ is the number of input channels, $H$ and $W$ denote the inline and crossline dimensions, and $D$ denotes the vertical time dimension. The volume is divided into nonoverlapping patches of size $\left(p_H, p_W, p_D\right)$, yielding a token grid of size $N_H \times N_W \times N_D$, where

$$N_H = \frac{H}{p_H}, N_W = \frac{W}{p_W}, N_D = \frac{D}{p_D}. \tag{1}$$

During pretraining, each patch is mapped to a token through factorized embedding and combined with direction-aware positional information. Trace-aligned tube masking removes vertically aligned token groups at selected lateral locations, and only the visible tokens are processed by the Transformer encoder. The decoder reconstructs the masked regions, after which a lightweight 3D convolutional refiner reduces local discontinuities near patch boundaries. The reconstruction branch is optimised using the seismic reconstruction objective described in Section 2.4.

For downstream inversion, masking is disabled and the encoder processes the complete token sequence. The encoded seismic features are combined with a low-frequency impedance trace in a trace-oriented prediction head to estimate the impedance at the center trace of each input window. Sliding-window inference is then used to assemble the predicted traces into impedance sections or volumes.

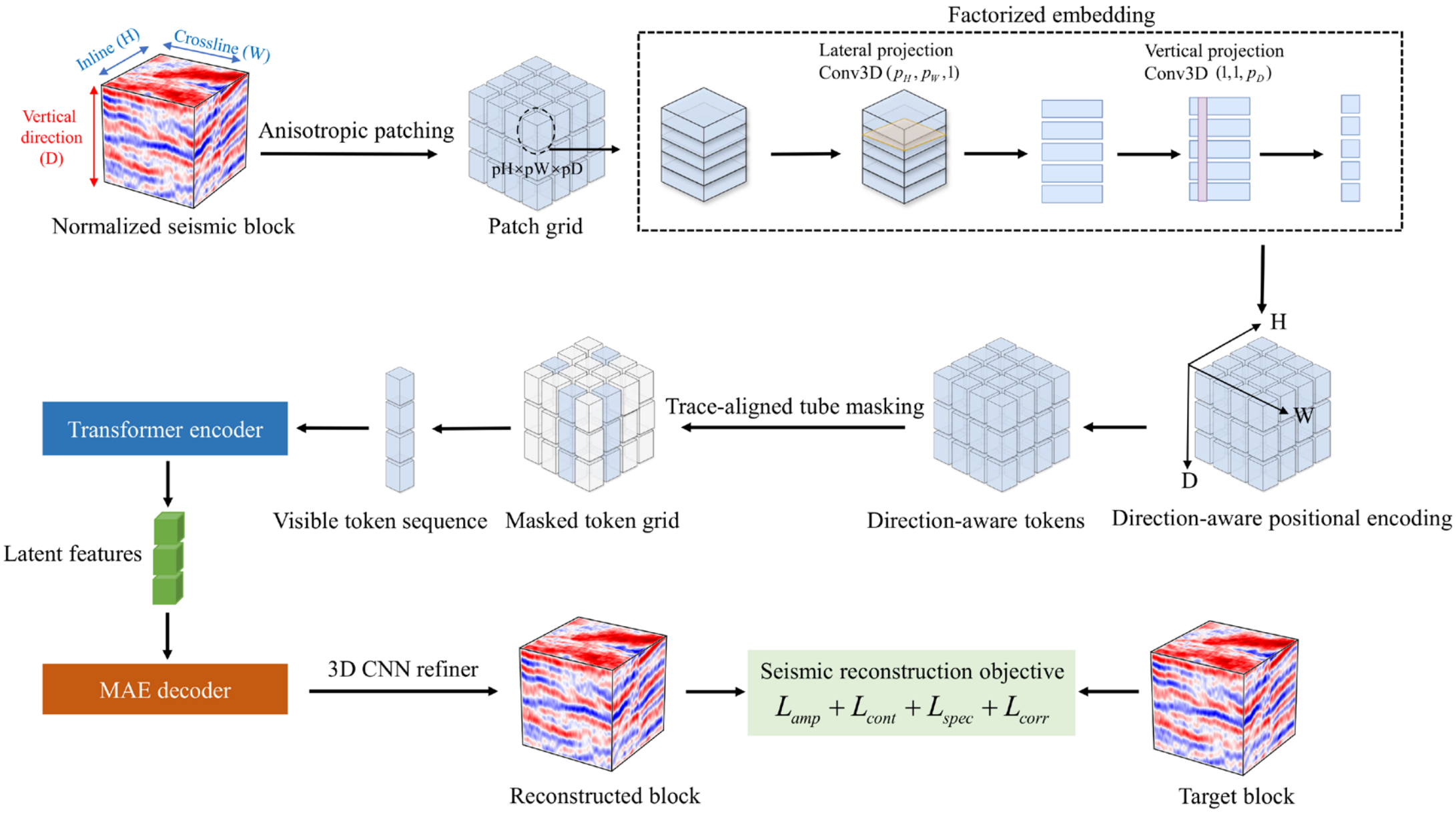

**Figure 1** Overview of the proposed DA-MAE self-supervised pretraining framework.

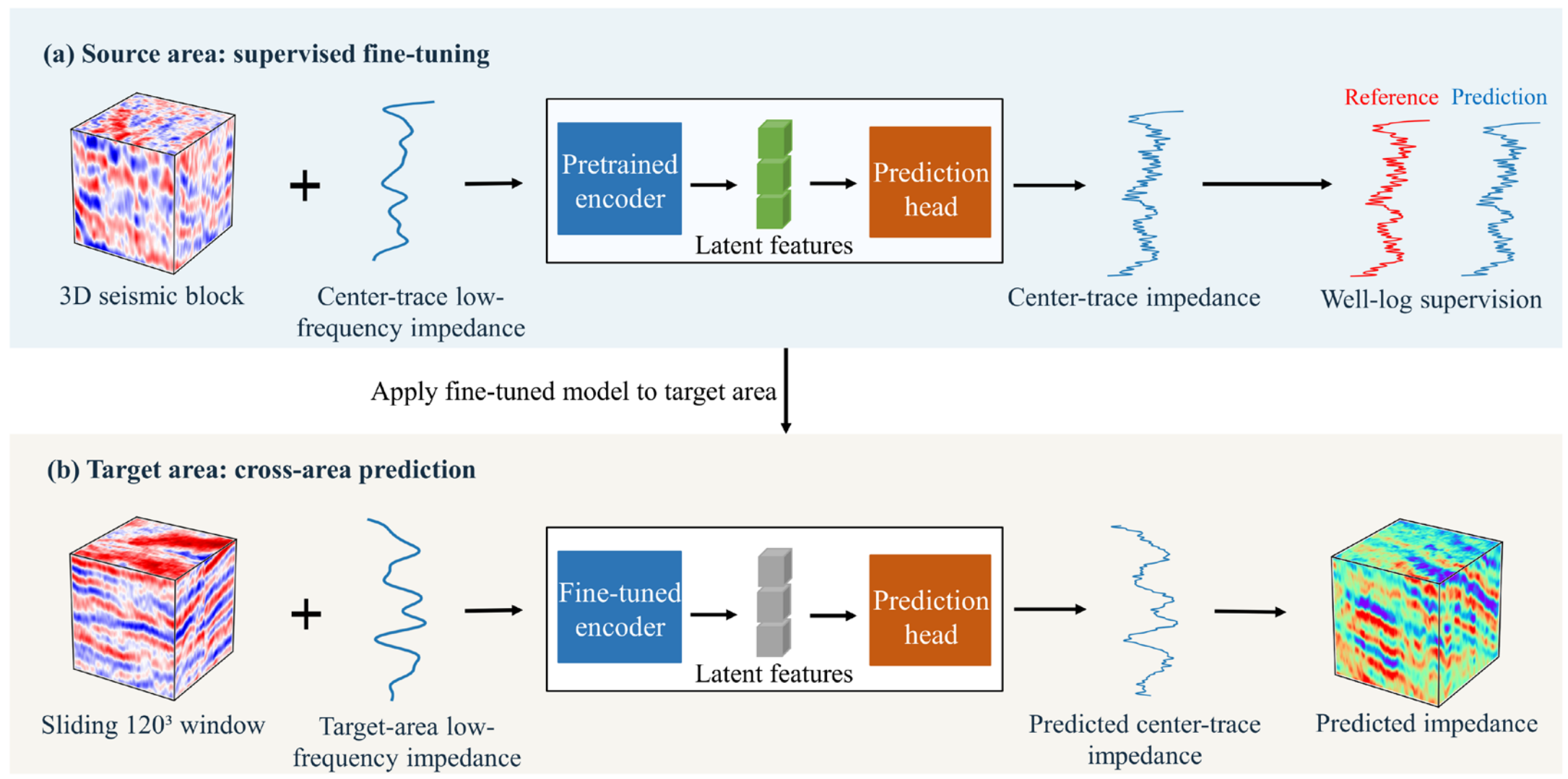

**Figure 2** Cross-area transfer of the pretrained DA-MAE encoder for acoustic impedance inversion.

## 2.2. Direction-Aware seismic token representation

DA-MAE represents 3D seismic data through anisotropic patching, factorized token embedding, and direction-aware positional encoding. These operations separately control lateral structural support and vertical waveform context while preserving positional information along the inline, crossline, and vertical directions.

### *2.2.1. Anisotropic patching*

The input volume is divided into nonoverlapping patches of size $(p_H, p_W, p_D)$. The lateral dimensions $(p_H, p_W)$ determine the spatial support used to represent reflector geometry and trace-to-trace variations, whereas $p_D$ determines the vertical waveform interval represented by each token. The three patch dimensions are selected independently rather than being constrained to the same size.

Larger lateral patches reduce the number of tokens but aggregate information over a broader spatial region, which may weaken the representation of local reflector geometry. Increasing $p_D$ incorporates a longer waveform interval into each token but reduces vertical resolution. Independent selection of $(p_H, p_W, p_D)$ therefore allows lateral structural resolution and vertical waveform context to be controlled separately.

### *2.2.2. Factorized token embedding*

Conventional 3D patch embedding projects the lateral and vertical dimensions in a single operation. To retain their different roles in seismic representation, DA-MAE factorizes the projection into lateral aggregation followed by vertical integration (Tran *et al*. 2018; Arnab *et al*. 2021):

$$Z = \text{Flatten}\left[E_D\left(E_{HW}\left(X\right)\right)\right]. \tag{2}$$

Here, $E_{HW}$ is a 3D convolution with kernel size and stride $(p_H, p_W, 1)$, which aggregates information over the lateral patch while preserving the vertical sample sequence. The subsequent projection $E_D$ uses kernel size and stride $(1, 1, p_D)$ to integrate the intermediate features over the specified vertical waveform interval. The intermediate channel dimension is controlled by a spatial embedding ratio.

The resulting feature map has dimensions $E \times N_H \times N_W \times N_D$ and is flattened into a token sequence $Z \in \mathrm{R}^{N \times E}$, where $N = N_H N_W N_D$. The factorized projection preserves the token grid while separating lateral aggregation from vertical integration.

*2.2.3. Direction-Aware positional encoding*

To preserve directional position information, DA-MAE uses separate trainable embeddings for the inline, crossline, and vertical coordinates (Wu *et al.* 2021). For a token located at $(i, j, k)$, the positional embedding is defined as

$$p_{i,j,k} = \left[ p_i^H \parallel p_j^W \parallel p_k^D \right], \tag{3}$$

where $\parallel$ denotes concatenation, and $p_i^H$, $p_j^W$ and $p_k^D$ denote the trainable positional embeddings for the inline, crossline, and vertical coordinates, respectively. Their dimensions satisfy $E_H + E_W + E_D = E$ and are allocated according to the ratio $(r_H, r_W, r_D)$. Separate positional embedding tables are used for the encoder and decoder.

## 2.3. Masking and reconstruction

The reconstruction stage includes trace-aligned tube masking, Transformer-based reconstruction, and local convolutional refinement. The masking strategy controls the context available to the encoder, while the decoder and refiner recover the missing seismic information.

*2.3.1. Trace-Aligned tube masking*

Conventional random masking removes tokens independently from the 3D token grid, so vertically adjacent tokens along the same trace may remain visible and provide strong local cues for reconstruction. To reduce this dependence on nearby information,

DA-MAE uses trace-aligned tube masking (Tong *et al.* 2022; Liu *et al.* 2024). A binary mask is first sampled on the lateral token grid $N_H \times N_W$ and then repeated along the vertical direction. For a lateral location $(i, j)$, the masking state is defined as

$$m_{i,j,k} = m_{i,j}, \quad k = 1, \ldots, N_D, \tag{4}$$

where $m_{i,j} = 1$ denotes a masked lateral location. Thus, all tokens along the same vertical sequence are masked or retained together. This removes the complete vertical token sequence at selected lateral locations and requires reconstruction to rely more on neighbouring traces and laterally coherent reflector structure. Only visible tokens are passed to the encoder. Random masking with the same masking ratio is used as a controlled baseline.

*2.3.2. Visible token encoding and reconstruction*

Each visible patch token is combined with its positional embedding and processed by the Transformer encoder. The encoded features are projected into the decoder embedding space, and learnable mask tokens are inserted at the missing positions to restore the original token grid. Decoder-specific positional embeddings are then added, and the decoder predicts the samples within each masked patch. The predicted patches are rearranged to form the reconstructed volume.

Because patchwise reconstruction may introduce local discontinuities near patch boundaries, a lightweight 3D convolutional refiner is applied to the decoder output. The completed seismic volume is defined as

$$\hat{X} = (1 - M) \odot X + M \odot \hat{X}_{\text{refined}}, \tag{5}$$

where $X$ denotes the original seismic volume, $\hat{X}$ denotes the completed seismic volume, and $\hat{X}_{\text{refined}}$ denotes the refined decoder output. $M$ is the voxel-level expansion of the token mask, with $M=1$ at masked locations and $M=0$ elsewhere, and $\odot$ denotes elementwise multiplication. Visible samples are therefore retained from the input, while masked samples are replaced by the refined predictions. The convolutional refiner is used only during pretraining and is not transferred to the downstream inversion model.

## 2.4. Seismic reconstruction objective

Pointwise amplitude loss alone does not explicitly account for reflector continuity or waveform characteristics. We therefore define a seismic reconstruction objective composed of four terms:

$$L_{\text{rec}}=\lambda_{\text{amp}}L_{\text{amp}}+\lambda_{\text{cont}}L_{\text{cont}}+\lambda_{\text{spec}}L_{\text{spec}}+\lambda_{\text{corr}}L_{\text{corr}}, \tag{6}$$

where $L_{\text{amp}}$, $L_{\text{cont}}$, $L_{\text{spec}}$ and $L_{\text{corr}}$ denote amplitude fidelity, directional structural continuity, spectral consistency, and trace-wise waveform similarity, respectively, with corresponding weights $\lambda_{\text{amp}}$, $\lambda_{\text{cont}}$, $\lambda_{\text{spec}}$, $\lambda_{\text{corr}}$. Let $X$ and $\hat{X}$ denote the target and completed seismic volumes, respectively.

### *2.4.1. Masked-Region amplitude fidelity*

Amplitude fidelity is enforced only within the masked regions using the Huber loss:

$$L_{\text{amp}}=\frac{\sum_{u}M_{u}\rho_{\delta}\left(\hat{X}_{u}\text{-}X_{u}\right)}{\sum_{u}M_{u}}, \tag{7}$$

where $M_u=1$ indicates a masked voxel, and $\rho_\delta$ is the Huber function with threshold $\delta$. This term constrains local amplitude errors while reducing sensitivity to large residuals.

*2.4.2. Directional structural continuity*

To preserve reflector continuity and directional seismic variations, we compare first-order gradients of the reconstructed and target volumes along the inline, crossline, and vertical directions:

$$L_{\text{cont}}=\sum_{q\in\{H,W,D\}}\alpha_q\left[\left\langle\left|\nabla_q\hat{X}-\nabla_q X\right|\right\rangle+\beta\left|\left\langle\left|\nabla_q\hat{X}\right|\right\rangle-\left\langle\left|\nabla_q X\right|\right\rangle\right|\right], \tag{8}$$

where $\nabla_q$ denotes the forward-difference operator along direction $q\in\{H,W,D\}$, $\alpha_q$ controls the contribution of each direction, $\beta$ weights the gradient-magnitude term, and $\langle\cdot\rangle$ denotes averaging over the volume. The first component matches local directional variations, while the second reduces systematic differences in overall gradient strength.

*2.4.3. Spectral consistency*

Spectral consistency is imposed by comparing the logarithmic amplitudes of the windowed 3D Fourier spectra:

$$\mathrm{L}_{\text{spec}}=\left\langle\left|\log\left(1+\left|\mathrm{F}_3\left(W\odot\hat{X}\right)\right|\right)-\log\left(1+\left|\mathrm{F}_3\left(W\odot X\right)\right|\right)\right|\right\rangle, \tag{9}$$

where $\mathrm{F}_3$ denotes the 3D Fourier transform and $W$ is a separable 3D Hann window. The window reduces boundary-related spectral leakage, while the logarithmic amplitude limits the dominance of high-energy components.

*2.4.4. Trace-Wise waveform correlation*

Waveform similarity within masked traces is further constrained using the Pearson correlation coefficient:

$$\mathrm{L}_{\text{corr}}=\frac{1}{\left|\mathrm{T}_M\right|}\sum_{\tau\in\mathrm{T}_M}\left[1\text{-PCC}\left(\hat{x}_{\tau,M},x_{\tau,M}\right)\right], \tag{10}$$

where $\mathrm{T}_M$ is the set of vertical traces containing masked samples, and $\hat{x}_{\tau,M}$ and $x_{\tau,M}$ denote the reconstructed and target samples within the masked portion of trace $\tau$, respectively. This term complements the amplitude loss by emphasizing waveform shape within masked trace segments.

### 2.5. Transfer to acoustic impedance inversion

Acoustic impedance inversion is used as the downstream task to evaluate the transferability of the learned seismic representations. For each target location, the model takes a centered 3D seismic window together with a low-frequency impedance trace. The low-frequency input provides the background impedance trend, while the seismic data constrain the higher-frequency impedance variations. The model predicts the impedance trace at the center of the input window, and sliding-window inference is used to form impedance sections or volumes.

During fine-tuning, masking is disabled and the complete seismic token sequence is processed by the pretrained encoder. The encoded seismic features are passed to a trace-oriented prediction head, which combines neighbouring features and vertical context to estimate the residual log-impedance trace. The final prediction is

$$\hat{z} = z_{\mathrm{LF}} + \Delta\hat{z}, \tag{11}$$

where $z_{\mathrm{LF}}$ is the low-frequency impedance trace and $\Delta\hat{z}$ is the impedance variation predicted from the seismic features. This separates the prescribed low-frequency background from the higher-frequency impedance information inferred from the seismic data.

Only the pretrained encoder is transferred; the MAE decoder and 3D convolutional refiner are discarded. The encoder and prediction head are jointly fine-tuned using labelled data from the source area and then applied to the target area without further parameter updates. All comparisons use the same prediction head, low-frequency background, target-area control information, preprocessing, inversion objective, and fine-tuning protocol.

## 3. Experimental setup

The experiments evaluate DA-MAE from two perspectives: masked seismic reconstruction and downstream representation transfer. Reconstruction performance is assessed on an independent field survey, while representation transfer is evaluated through cross-area acoustic impedance inversion. This section describes the datasets, training settings, and evaluation protocols.

### 3.1. Datasets, preprocessing, and data partitioning

#### *3.1.1. Pretraining and external reconstruction datasets*

The pretraining dataset comprises 33 3D post-stack seismic surveys from China, the Gulf of Mexico, the Netherlands, Congo, and Uganda, covering diverse geological settings and seismic characteristics. All volumes were resampled to a temporal sampling interval of 1 ms and divided into nonoverlapping 120×120×120 blocks. To reduce amplitude differences among surveys, each seismic volume was normalised using its 99.5th percentile absolute amplitude:

$$a_s = Q_{0.995}\left(\left|X_s\right|\right),\ \bar{X}_s = \frac{X_s}{a_s + \epsilon}, \tag{12}$$

where $X_s$ denotes the seismic volume from survey $s$, $Q_{0.995}$ denotes the 99.5th percentile, and $\epsilon$ is a small constant for numerical stability.

The resulting dataset contains 35,718 blocks, including 32,146 for pretraining and 3,572 for validation and checkpoint selection. The split was performed at the survey level to avoid data leakage. An independent field survey from eastern China, containing 380 blocks, was reserved exclusively for external reconstruction evaluation.

### *3.1.2. Cross-Area acoustic impedance inversion datasets*

Cross-area representation transfer was evaluated using two field surveys. Area C served as the source area for supervised fine-tuning, whereas Area L served as the target area for validation and testing. Their seismic volumes contain 577×1441×765 and 481×385×1046 samples, respectively. Area C contains 20 wells with paired seismic and impedance logs, which provided all labelled data for downstream fine-tuning.

Area L contains ten wells, denoted Wells A–J. Wells A and B were used for validation and checkpoint selection and to construct the 8-Hz low-frequency impedance background and wavelet information. Wells C–J were reserved for test-well evaluation. Their impedance logs were excluded from model training, checkpoint selection, low-frequency model construction, and wavelet estimation, and were used only for final quantitative evaluation.

The seismic data from Area L were included in self-supervised pretraining without impedance labels, but no target-area impedance labels were used for gradient-based downstream fine-tuning. The experiment therefore represents cross-area transfer with limited target-area control rather than fully target-blind transfer. The same target-area control information and preprocessing settings were used for all downstream comparisons. Table 1 summarizes the datasets and their roles in the experiments.

**Table 1** Summary of the seismic datasets used in this study.

| Dataset | Data Scale | Wells/Surveys | Experimental role |
|---|---|---|---|
| Pretraining dataset | 35,718 blocks | 33 surveys | Self-supervised training and internal validation |
| External test dataset | 380 blocks | 1 independent survey | Reconstruction generalisation evaluation |

| Dataset | Data Scale | Wells/Surveys | Experimental role |
|---|---|---|---|
| Area C | 577×1441×765 samples | 20 wells | Supervised downstream fine-tuning |
| Area L | 481×385×1046 samples | 10 wells | 2 validation wells and 8 test wells |

## 3.2. Implementation and training settings

### *3.2.1. DA-MAE pretraining*

DA-MAE was implemented in PyTorch and pretrained on a single NVIDIA RTX 6000 GPU. Unless otherwise specified, the encoder contains 16 Transformer blocks with an embedding dimension of 768 and 12 attention heads, while the decoder contains 6 Transformer blocks with an embedding dimension of 576 and 8 heads. The masking ratio was fixed at 75%. The full DA-MAE includes factorized token embedding, direction-aware positional encoding, trace-aligned tube masking, and the convolutional refiner.

All models were pretrained for 100 epochs using AdamW with an initial learning rate of $1\times10^{-4}$ , a weight decay of 0.05, and cosine decay to $1\times10^{-6}$. Mixed-precision training was used. The physical batch size and gradient-accumulation steps were adjusted across token configurations to maintain an effective batch size of 64. For each configuration, the checkpoint with the lowest validation loss was selected for external reconstruction evaluation and downstream transfer.

For the seismic reconstruction objective, the Huber threshold was set to $\delta = 0.1$ ,with $\lambda_{\text{amp}} = 1.0$ , $\lambda_{\text{cont}} = 0.5$ , $\lambda_{\text{spec}} = 0.02$ and $\lambda_{\text{corr}} = 0.5$ . The directional weights were $\alpha_H = 1.0$ , $\alpha_W = 1.0$ , and $\alpha_D = 1.5$ , with $\beta = 0.5$ . These settings were kept fixed unless otherwise stated.

*3.2.2. Downstream inversion and fine-tuning*

For acoustic impedance inversion, the trace-oriented prediction head contains 6 Transformer decoder blocks with an embedding dimension of 576 and 8 attention heads, followed by a lightweight 1D convolutional prediction head. The pretrained encoder and prediction head were jointly fine-tuned for 20 epochs using AdamW with a learning rate of $1\times10^{-4}$, a weight decay of $1\times10^{-4}$, and a batch size of 4.

Each input consists of a 120×120×120 seismic window. Training windows were sampled with a vertical stride of 5 samples, while validation and test inference used a stride of 1 sample. Predictions from overlapping windows were averaged to reconstruct complete impedance traces, which were then assembled laterally into impedance sections or volumes.

Seismic amplitudes were normalised using quantile-based scaling, while acoustic impedance was transformed to the log domain and standardised using statistics from the source-area training data. An 8-Hz impedance background was used as the low-frequency prior. Wavelet information was included as an additional conditioning input using the estimated dominant frequency and phase. For Area L, both the low-frequency background and wavelet parameters were derived only from the target-area control wells. The same conditioning procedure was used for all downstream models.

The downstream objective combines multiscale impedance reconstruction, phase consistency, vertical-gradient matching, band-pass fidelity, and seismic forward consistency. The detailed formulations and weighting parameters are provided in Appendix A.2.

## 3.3. Evaluation and ablation protocol

Reconstruction performance was evaluated using mean squared error (MSE), 3D structural similarity (SSIM), Pearson correlation coefficient (PCC), spectral fidelity (SF), phase fidelity (PF), lateral gradient error (LGE), and vertical gradient error (TGE). MSE measures amplitude reconstruction error; SSIM and PCC assess structural and waveform similarity; SF and PF characterise spectral and phase agreement; and LGE and TGE quantify reconstruction errors in lateral reflector continuity and vertical waveform variation, respectively. Higher SSIM, PCC, SF, and PF and lower MSE, LGE, and TGE indicate better reconstruction. Complete metric definitions are provided in Appendix A.3.

All reconstruction metrics were averaged over the 380 blocks from the independent external survey. For each token configuration and masking strategy, the same evaluation masks were used to ensure consistent comparison. Random and tube masking were evaluated at the same masking ratio.

Acoustic impedance prediction was evaluated using normalised mean absolute error (NMAE), normalised root-mean-squared error (NRMSE), and PCC. For a reference trace containing $N$ valid samples,

$$\text{NMAE}=\frac{1}{N\left(z_{\max}-z_{\min}\right)}\sum_{i=1}^{N}\left|\hat{z}_i-z_i\right|, \tag{13}$$

$$\text{NRMSE}=\frac{1}{z_{\max}-z_{\min}}\sqrt{\frac{1}{N}\sum_{i=1}^{N}\left(\hat{z}_i-z_i\right)^2}. \tag{14}$$

Here, $\hat{z}_i$ and $z_i$ denote the predicted and reference impedance values, respectively, and $z_{\max}-z_{\min}$ is the amplitude range of the reference trace. The metrics were computed separately for each test well and then averaged over Wells C–J.

For interwell evaluation, overlapping center-trace predictions were averaged and assembled along selected profiles to form impedance sections. The resulting sections were examined for lateral continuity and consistency with the seismic structure and available well logs. Controlled experiments were conducted to examine the main factors affecting reconstruction and representation transfer. The conventional 3D MAE Backbone (B), which uses a single 3D patch projection and standard learnable positional encoding, serves as the reference architecture. Unless otherwise stated, only the factor under investigation is varied within each controlled comparison, while the data partition, training settings, and evaluation procedure are kept unchanged.

## 4. Reconstruction results and analysis

### 4.1. Effect of the seismic reconstruction objective

The seismic reconstruction objective was compared with amplitude-only training using the same Backbone architecture and training settings. As shown in Table 2, the seismic objective improved all reconstruction metrics across the three representative token configurations, with the largest gains observed under fine tokenization. For 4×4×8 patches, MSE decreased from $4.58\times10^{-2}$ to $6.14\times10^{-3}$, while SSIM increased from 0.287 to 0.757. The improvement was smaller for the coarser 8×8×30 configuration, where MSE decreased from $1.73\times10^{-2}$ to $1.32\times10^{-2}$.

The representative reconstructions in Fig. 3 show the same trend. With 4×4×8 tokenization, amplitude-only training shows weaker recovery of coherent reflection events and more pronounced reflector-aligned residuals. In contrast, the seismic reconstruction objective produces more continuous reflections and weaker residuals. The visual difference is smaller for 8×8×30 patches, consistent with the more modest quantitative improvement under coarser tokenization.

These results indicate that amplitude-only training is less effective at preserving reflector structure and waveform characteristics, particularly under fine tokenization. The seismic reconstruction objective is therefore used in all subsequent reconstruction experiments.

**Table 2** Ablation study of the seismic reconstruction objective on representative patch configurations.

| Patch size | Loss | MSE | SSIM | PCC | PF | SF | LGE | TGE |
|---|---|---|---|---|---|---|---|---|
| 4×4×8 | Amplitude-only | $4.58\times10^{-2}$ | 0.287 | 0.548 | 0.832 | 0.735 | $4.36\times10^{-2}$ | $2.20\times10^{-2}$ |
| 4×4×8 | Seismic | $6.14\times10^{-3}$ | 0.757 | 0.951 | 0.961 | 0.977 | $2.49\times10^{-2}$ | $1.01\times10^{-2}$ |
| 8×8×15 | Amplitude-only | $4.59\times10^{-2}$ | 0.304 | 0.547 | 0.802 | 0.785 | $3.47\times10^{-2}$ | $2.22\times10^{-2}$ |
| 8×8×15 | Seismic | $1.33\times10^{-2}$ | 0.634 | 0.895 | 0.930 | 0.952 | $2.77\times10^{-2}$ | $1.30\times10^{-2}$ |
| 8×8×30 | Amplitude-only | $1.73\times10^{-2}$ | 0.545 | 0.853 | 0.913 | 0.938 | $3.28\times10^{-2}$ | $1.78\times10^{-2}$ |
| 8×8×30 | Seismic | $1.32\times10^{-2}$ | 0.641 | 0.896 | 0.930 | 0.953 | $2.68\times10^{-2}$ | $1.28\times10^{-2}$ |

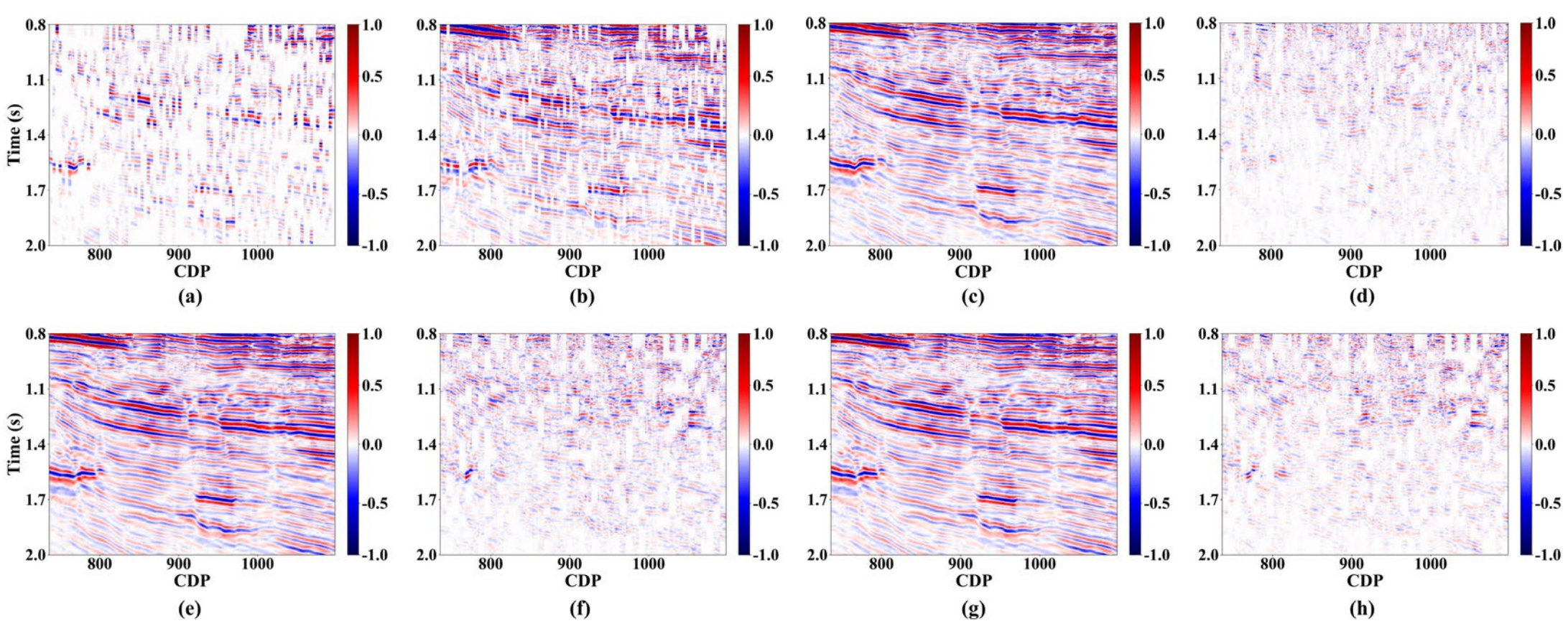

**Figure 3** Reconstruction results and corresponding residual maps obtained using the amplitude-only baseline and the proposed seismic reconstruction objective. Panels (a)–(d) and (e)–(h) correspond to patch sizes of 4×4×8 and 8×8×30, respectively. Within each group, the first two panels show the amplitude-only reconstruction and its residual, followed by the seismic reconstruction and its residual. Amplitude-only denotes training using $L_{\text{amp}}$ alone.

## 4.2. Effect of token scale and embedding capacity

Token geometry determines both the local seismic support represented by each token and the resulting sequence length. For a 120×120×120 input volume, 4×4×8 patches yield 13,500 tokens before masking, whereas 8×8×30 patches yield only 900, illustrating the tradeoff between token resolution and computational cost. We first varied the lateral and vertical patch dimensions independently using the same Backbone, and then examined coupled changes in patch size and embedding capacity.

*4.2.1. Backbone-Level patch-size sensitivity*

Two patch-size sweeps were conducted using the same Backbone and seismic reconstruction objective. In the first, the vertical patch length was varied from 6 to 120 samples while the lateral size was fixed at 8×8. In the second, the lateral patch size was varied from 4×4 to 12×12 while the vertical length was fixed at 8 samples.

Reconstruction performance remained relatively stable across moderate vertical patch lengths (Table 3a). Configurations from 8×8×6 to 8×8×30 produced comparable results: 8×8×10 yielded the lowest MSE, while 8×8×8 performed slightly better on most other metrics. More noticeable degradation appeared only at vertical lengths of 60 and 120 samples. This indicates that reconstruction is relatively insensitive to moderate changes in vertical token extent, whereas substantially longer vertical patches reduce reconstruction fidelity.

Lateral token scale had a much stronger effect (Table 3b). Increasing the lateral patch size from 4×4 to 12×12 increased MSE from $6.14\times10^{-3}$ to $1.86\times10^{-2}$, while SSIM and PCC decreased from 0.757 and 0.951 to 0.574 and 0.847, respectively. Figure 4 also shows increasingly pronounced reflector-aligned residuals as the lateral patches become larger. These results show that reconstruction is substantially more sensitive to lateral token resolution than to moderate changes in vertical patch length. The 4×4×8 configuration achieved the highest reconstruction fidelity, although it also produced the longest token sequence.

**Table 3a** Reconstruction sensitivity to vertical patch length with a fixed lateral patch size of 8×8 under a fixed backbone.

| Patch size | MSE | SSIM | PCC | PF | SF | LGE | TGE |
|---|---|---|---|---|---|---|---|
| 8×8×6 | $1.33\times10^{-2}$ | 0.639 | 0.896 | 0.930 | 0.953 | $2.67\times10^{-2}$ | $1.30\times10^{-2}$ |
| 8×8×8 | $1.27\times10^{-2}$ | 0.651 | 0.900 | 0.932 | 0.955 | $2.57\times10^{-2}$ | $1.25\times10^{-2}$ |
| 8×8×10 | $1.26\times10^{-2}$ | 0.650 | 0.899 | 0.931 | 0.954 | $2.59\times10^{-2}$ | $1.27\times10^{-2}$ |
| 8×8×15 | $1.33\times10^{-2}$ | 0.634 | 0.895 | 0.930 | 0.952 | $2.77\times10^{-2}$ | $1.30\times10^{-2}$ |
| 8×8×20 | $1.30\times10^{-2}$ | 0.645 | 0.898 | 0.931 | 0.954 | $2.64\times10^{-2}$ | $1.27\times10^{-2}$ |
| 8×8×30 | $1.32\times10^{-2}$ | 0.641 | 0.896 | 0.930 | 0.953 | $2.68\times10^{-2}$ | $1.28\times10^{-2}$ |
| 8×8×60 | $1.37\times10^{-2}$ | 0.630 | 0.893 | 0.929 | 0.951 | $2.81\times10^{-2}$ | $1.29\times10^{-2}$ |
| 8×8×120 | $1.43\times10^{-2}$ | 0.614 | 0.887 | 0.928 | 0.949 | $2.97\times10^{-2}$ | $1.32\times10^{-2}$ |

**Table 3b** Reconstruction sensitivity to lateral patch size with a fixed vertical patch length of 8 samples under a fixed backbone.

| Patch size | MSE | SSIM | PCC | PF | SF | LGE | TGE |
|---|---|---|---|---|---|---|---|
| 4×4×8 | $6.14\times10^{-3}$ | 0.757 | 0.951 | 0.961 | 0.977 | $2.49\times10^{-2}$ | $1.01\times10^{-2}$ |
| 5×5×8 | $6.81\times10^{-3}$ | 0.746 | 0.946 | 0.957 | 0.975 | $2.34\times10^{-2}$ | $9.67\times10^{-3}$ |
| 6×6×8 | $8.31\times10^{-3}$ | 0.716 | 0.934 | 0.950 | 0.969 | $2.39\times10^{-2}$ | $1.04\times10^{-2}$ |
| 8×8×8 | $1.27\times10^{-2}$ | 0.651 | 0.900 | 0.932 | 0.955 | $2.57\times10^{-2}$ | $1.25\times10^{-2}$ |
| 10×10×8 | $1.50\times10^{-2}$ | 0.615 | 0.877 | 0.922 | 0.943 | $2.77\times10^{-2}$ | $1.35\times10^{-2}$ |
| 12×12×8 | $1.86\times10^{-2}$ | 0.574 | 0.847 | 0.907 | 0.931 | $2.86\times10^{-2}$ | $1.47\times10^{-2}$ |

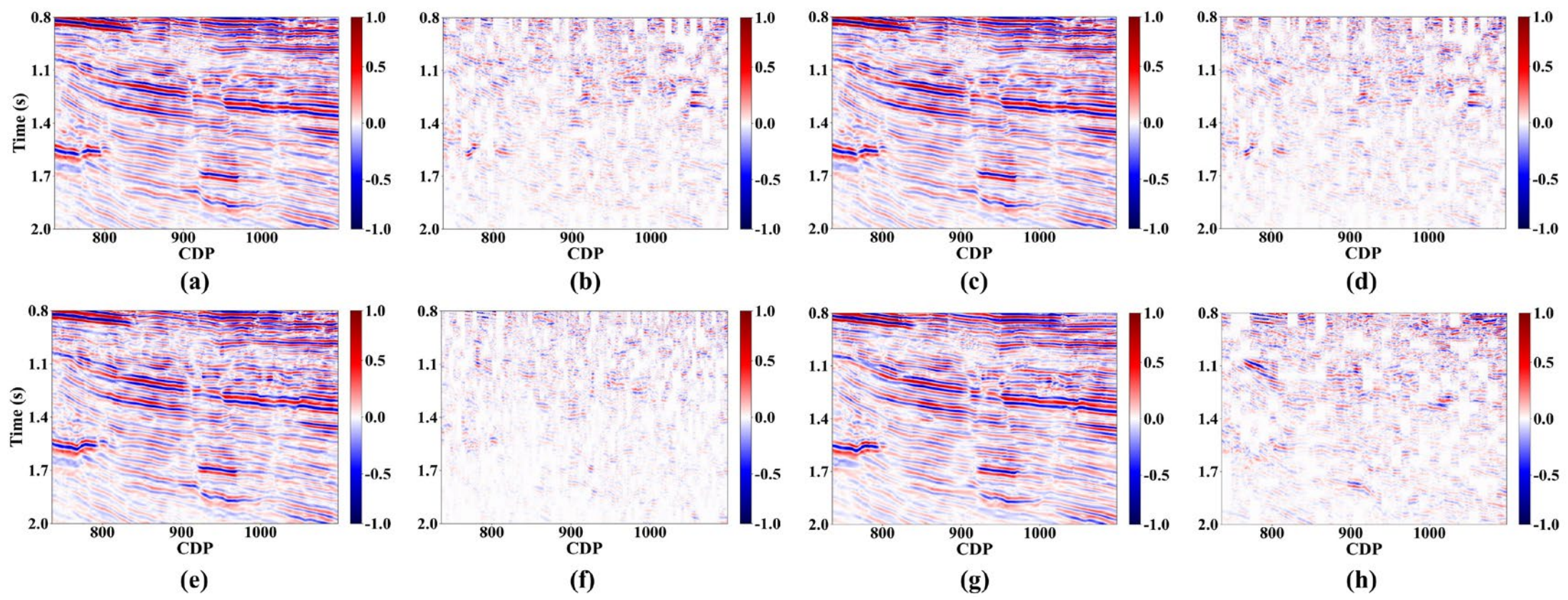

**Figure 4** Representative reconstruction results and residual maps under different token scales. Panels (a)–(b), (c)–(d), (e)–(f) and (g)–(h) correspond to patch sizes of 8×8×6, 8×8×30, 4×4×8 and 12×12×8, respectively, with the reconstruction followed by the corresponding residual in each pair.

*4.2.2. Coupled patch–embedding scaling*

We next examined whether increasing encoder capacity could compensate for the loss of local information associated with coarser tokenization. Three full DA-MAE configurations were evaluated. C1 used 4×4×8 patches with an embedding dimension of 512 and 8 attention heads; C2 used 6×6×12 patches with an embedding dimension of 768 and 12 heads; and C3 used 8×8×15 patches with an embedding dimension of 1024 and 16 heads. The dimension of each attention head was fixed at 64, and all other settings were kept unchanged.

Despite the increase in encoder capacity, reconstruction performance declined as tokenization became coarser (Table 4). From C1 to C3, MSE increased from $6.20\times10^{-3}$ to $1.30\times10^{-2}$, while SSIM and PCC decreased from 0.771 and 0.954 to 0.651 and 0.899, respectively. Figure 5 also shows a progressive loss of local reflection detail and stronger structured residuals. Within the evaluated scaling schedule, increasing the embedding dimension and number of attention heads did not compensate for the degradation associated with coarser tokenization. These results indicate that token granularity and encoder capacity play distinct roles and are not directly interchangeable.

**Table 4** Quantitative results of the coupled patch-embedding scaling experiment.

| Configuration | MSE | SSIM | PCC | PF | SF | LGE | TGE |
|---|---|---|---|---|---|---|---|
| C1 | $6.20\times10^{-3}$ | 0.771 | 0.954 | 0.961 | 0.979 | $2.24\times10^{-2}$ | $8.70\times10^{-3}$ |
| C2 | $8.40\times10^{-3}$ | 0.719 | 0.934 | 0.950 | 0.970 | $2.36\times10^{-2}$ | $1.01\times10^{-2}$ |
| C3 | $1.30\times10^{-2}$ | 0.651 | 0.899 | 0.931 | 0.953 | $2.52\times10^{-2}$ | $1.21\times10^{-2}$ |

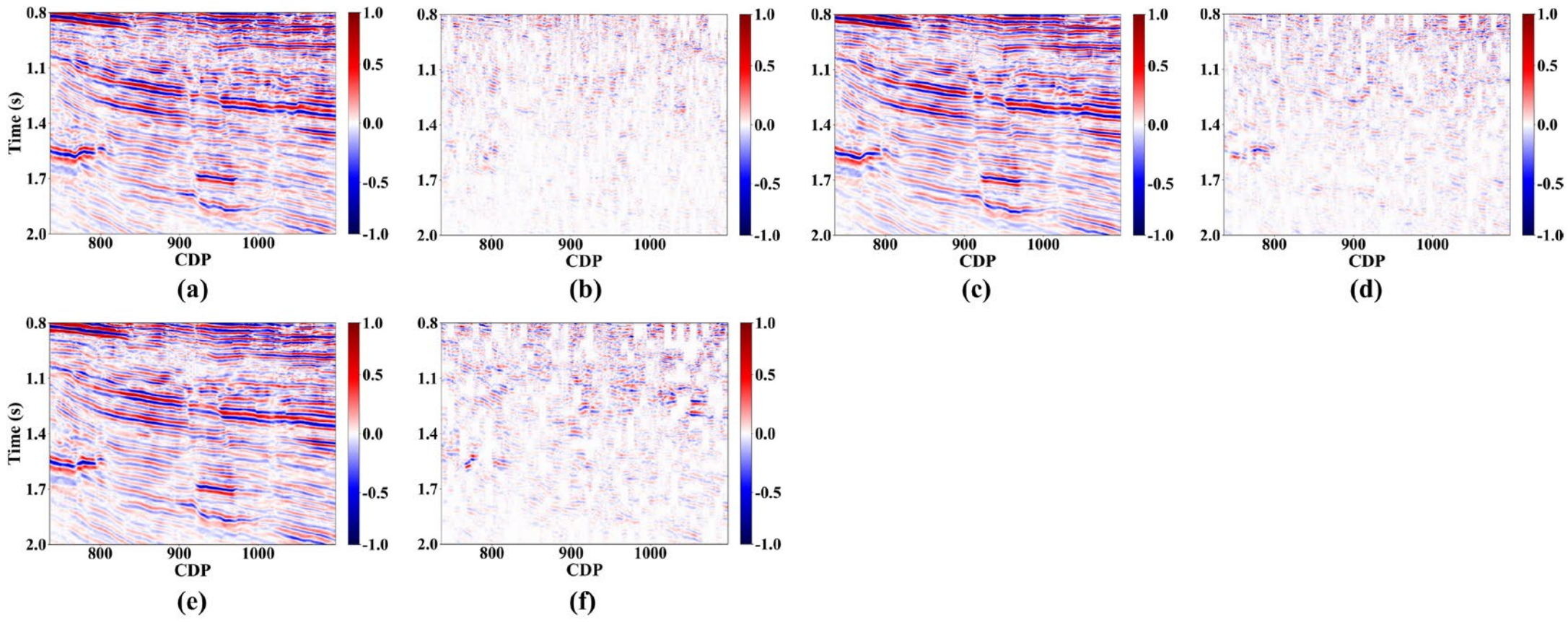

**Figure 5** Reconstruction results and corresponding residual maps for the coupled patch-embedding scaling experiment. Panels (a)–(b), (c)–(d), and (e)–(f) correspond to configurations C1, C2, and C3, respectively, with the reconstruction followed by the corresponding residual in each pair.

## 4.3. Progressive architectural component analysis

To examine the incremental effects of the architectural components, we started from the MAE Backbone (B), which uses a single 3D patch projection and a standard learnable positional embedding. The decoder-side convolutional refiner was then added to form B + Ref, followed by factorized token embedding to form B + Ref + FTE, and finally direction-aware positional encoding to obtain the Full model. The seismic reconstruction objective, tube-masking strategy, masking ratio, and training protocol were kept fixed. Because the components were introduced sequentially, these comparisons reflect fixed-order conditional effects rather than independent contributions of individual components.

For the 4×4×8 configuration, adding the convolutional refiner left MSE nearly unchanged but improved SSIM and reduced both gradient errors (Table 5). The subsequent addition of factorized token embedding produced the clearest improvement, yielding the lowest MSE and the best PCC, LGE, and TGE. Adding direction-aware positional encoding produced only minor changes in reconstruction performance, with the Full model remaining comparable to B + Ref + FTE on most metrics. The differences were smaller with the coarser 8×8×30 tokenization. Adding the refiner produced a modest reduction in MSE, whereas the subsequent additions of factorized token embedding and direction-aware positional encoding resulted in only minor changes. The Full model achieved the highest SSIM, PF, and SF, but most metrics remained close to those of B + Ref + FTE.

Overall, factorized token embedding provided the clearest incremental benefit under fine tokenization, while the additional reconstruction gain from direction-aware positional encoding was limited in this fixed-order comparison. Its contribution to downstream transfer is examined separately in Section 5.

**Table 5** Progressive architectural component analysis under 4×4×8 and 8×8×30 tokenization.

| Patch Size | Variant | MSE | SSIM | PCC | PF | SF | LGE | TGE |
|---|---|---|---|---|---|---|---|---|
| 4×4×8 | B | $6.14\times10^{-3}$ | 0.757 | 0.951 | 0.961 | 0.977 | $2.49\times10^{-2}$ | $1.01\times10^{-2}$ |
| 4×4×8 | B + Ref | $6.16\times10^{-3}$ | 0.773 | 0.954 | 0.961 | 0.979 | $2.24\times10^{-2}$ | $8.72\times10^{-3}$ |
| 4×4×8 | B + Ref + FTE | $5.92\times10^{-3}$ | 0.775 | 0.955 | 0.961 | 0.979 | $2.19\times10^{-2}$ | $8.52\times10^{-3}$ |
| 4×4×8 | Full | $6.15\times10^{-3}$ | 0.770 | 0.953 | 0.960 | 0.979 | $2.23\times10^{-2}$ | $8.65\times10^{-3}$ |
| 8×8×30 | B | $1.32\times10^{-2}$ | 0.641 | 0.896 | 0.930 | 0.953 | $2.68\times10^{-2}$ | $1.28\times10^{-2}$ |
| 8×8×30 | B + Ref | $1.27\times10^{-2}$ | 0.649 | 0.900 | 0.932 | 0.954 | $2.59\times10^{-2}$ | $1.21\times10^{-2}$ |
| 8×8×30 | B + Ref + FTE | $1.27\times10^{-2}$ | 0.652 | 0.901 | 0.932 | 0.954 | $2.57\times10^{-2}$ | $1.21\times10^{-2}$ |
| 8×8×30 | Full | $1.27\times10^{-2}$ | 0.653 | 0.901 | 0.933 | 0.956 | $2.57\times10^{-2}$ | $1.21\times10^{-2}$ |

*Note:* B denotes the MAE backbone. Ref and FTE denote the sequential addition of the decoder-side convolutional refiner and factorized token embedding, respectively. Full denotes B+Ref+FTE with direction-aware positional encoding. The reported results therefore reflect fixed-order incremental effects.

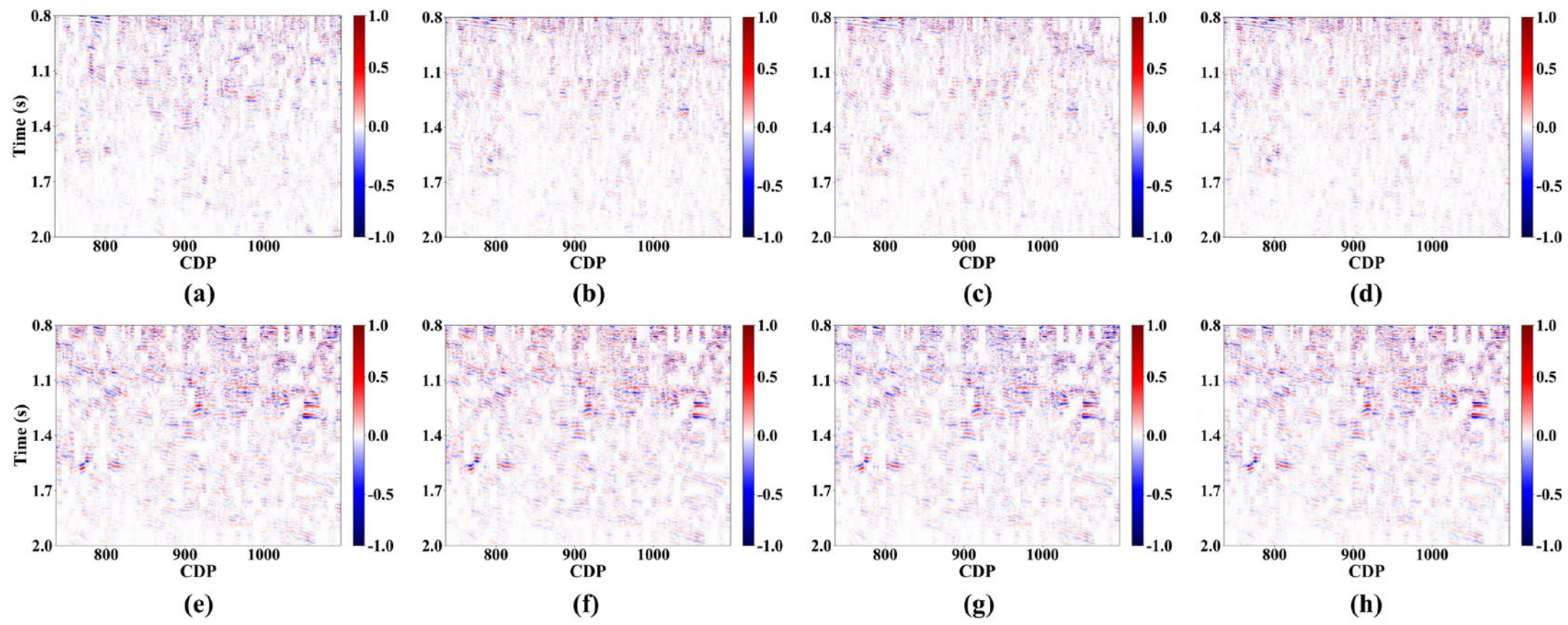

**Figure 6** Residual maps from the progressive architectural component analysis. Columns correspond to B, B+Ref, B+Ref+FTE, and Full, while rows correspond to the 4×4×8 and 8×8×30 tokenization settings.

## 4.4. Effect of masking strategy

To isolate the effect of masking geometry, random masking and trace-aligned tube masking were compared using the same patch size, architecture, reconstruction objective, masking ratio, and training protocol. Random masking achieved better reconstruction performance at both token scales (Table 6). For 4×4×8 patches, MSE decreased from $6.15\times10^{-3}$ with tube masking to $2.10\times10^{-3}$ with random masking, while SSIM increased from 0.770 to 0.897. Similar improvements were observed for 8×8×30 patches.

Figure 7 shows the same pattern, with smaller residuals under random masking. Because randomly masked tokens are often surrounded by visible neighbours, their reconstruction can rely more strongly on local context. Tube masking removes the entire vertical token sequence at selected lateral locations and therefore requires greater use of neighbouring traces and laterally coherent structures, increasing the contextual demand of reconstruction.

Therefore, the lower reconstruction error achieved by random masking does not necessarily imply better representation transferability. The downstream effects of the two masking strategies are evaluated directly in Section 5.1.

**Table 6** Reconstruction performance obtained using random and tube masking.

| Patch Size | Masking | MSE | SSIM | PCC | PF | SF | LGE | TGE |
|---|---|---|---|---|---|---|---|---|
| 8×8×30 | Random | $8.11\times10^{-3}$ | 0.733 | 0.940 | 0.953 | 0.973 | $2.28\times10^{-2}$ | $1.06\times10^{-2}$ |
| 8×8×30 | Tube | $1.27\times10^{-2}$ | 0.653 | 0.901 | 0.933 | 0.956 | $2.57\times10^{-2}$ | $1.21\times10^{-2}$ |
| 4×4×8 | Random | $2.10\times10^{-3}$ | 0.897 | 0.984 | 0.981 | 0.992 | $1.70\times10^{-2}$ | $6.88\times10^{-3}$ |
| 4×4×8 | Tube | $6.15\times10^{-3}$ | 0.770 | 0.953 | 0.960 | 0.979 | $2.23\times10^{-2}$ | $8.65\times10^{-3}$ |

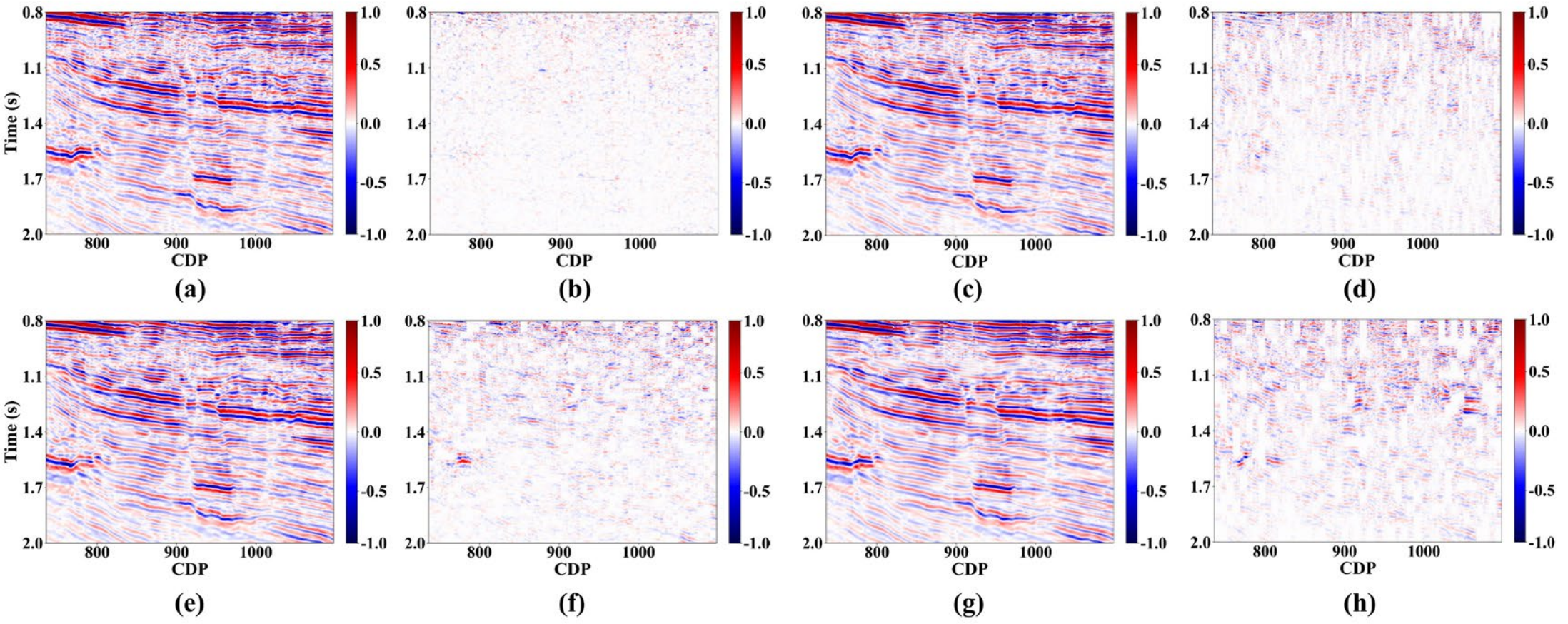

**Figure 7** Reconstruction results and corresponding residual maps for different masking strategies. Panels (a)–(b), (c)–(d), (e)–(f), and (g)–(h) correspond to 4×4×8 with random masking, 4×4×8 with tube masking, 8×8×30 with random masking, and 8×8×30 with tube masking, respectively, with the reconstruction followed by the corresponding residual in each pair.

## 5. Cross-Area representation transfer via acoustic impedance inversion

This section evaluates the transferability of the pretrained seismic representations through cross-area acoustic impedance inversion. Acoustic impedance provides a quantitative description of subsurface elastic contrasts and is commonly used to characterise stratigraphic variation and support reservoir interpretation. The models are fine-tuned using seismic–impedance pairs from Area C and then evaluated in Area L under the protocol described in Section 3.1.2.

### 5.1. Well-Level transfer and controlled comparisons

Under the matched Full architecture with 8×8×30 patches, self-supervised pretraining consistently improved cross-area impedance prediction relative to random initialization (Table 7). The pretrained model reduced NRMSE from $6.81\times10^{-2}$ to $5.47\times10^{-2}$ and produced impedance variations that more closely followed the reference logs. This confirms that the pretrained seismic representations retain useful information for impedance prediction when transferred to a different field area.

Under the finer 4×4×8 tokenization, the Full DA-MAE further improved the cross-area inversion relative to the pretrained Backbone, reducing NRMSE by 20.8% under matched downstream settings. As shown in Fig. 8, the predicted impedance traces reproduce both the regional background trend and many of the shorter-scale variations observed in the reference logs. The agreement across the eight test wells indicates that the transferred representations preserve seismic information relevant to local impedance contrasts rather than only the smooth low-frequency background.

The Backbone-level experiments further show that improved reconstruction does not necessarily lead to improved impedance inversion. In particular, the token configuration favoured by reconstruction was not consistently preferred for the downstream task. Finer lateral tokenization generally benefited inversion, suggesting that preserving local reflector geometry and trace-to-trace variations is important for recovering lateral impedance changes. In contrast, a broader vertical context can remain useful for inversion even when it produces lower reconstruction fidelity. Complete Backbone-level results are provided in Table A1.

The effect of masking geometry is likewise task dependent. Random masking produces higher reconstruction fidelity, whereas trace-aligned tube masking performs better for inversion under fine tokenization. By removing complete vertical token sequences at selected lateral locations, tube masking places greater emphasis on information from neighbouring traces and laterally coherent reflectors. This broader cross-trace information may be more useful for impedance inversion than the short-range context that favours masked reconstruction.

**Table 7** Controlled cross-area impedance-inversion results for different encoder initializations, architectures, and masking strategies.

| Encoder / initialization | Encoder architecture | Patch size | Masking | NMAE | NRMSE | PCC |
|---|---|---|---|---|---|---|
| Low-frequency baseline | - | - | - | $5.47\times10^{-2}$ | $7.44\times10^{-2}$ | 0.945 |
| Random | Full | 8×8×30 | - | $4.80\times10^{-2}$ | $6.81\times10^{-2}$ | 0.960 |
| Pretrained | Backbone | 8×8×30 | Tube | $4.60\times10^{-2}$ | $6.45\times10^{-2}$ | 0.966 |
| Pretrained | Full | 8×8×30 | Random | $4.00\times10^{-2}$ | $5.48\times10^{-2}$ | 0.975 |
| Pretrained | Full | 8×8×30 | Tube | $3.97\times10^{-2}$ | $5.47\times10^{-2}$ | 0.975 |
| Pretrained | Backbone | 4×4×8 | Tube | $4.45\times10^{-2}$ | $6.15\times10^{-2}$ | 0.968 |
| Pretrained | Full | 4×4×8 | Random | $3.99\times10^{-2}$ | $5.67\times10^{-2}$ | 0.972 |
| Pretrained | Full | 4×4×8 | Tube | $3.49\times10^{-2}$ | $4.87\times10^{-2}$ | 0.979 |

*Note:* Random initialization denotes the Full encoder architecture trained without self-supervised pretraining. Backbone and Full denote the baseline and complete encoder architectures, respectively. All pretrained Full models use the seismic reconstruction objective. The low-frequency baseline consists solely of the 8-Hz background model.

**Table 8** Well-level acoustic impedance prediction using the full DA-MAE encoder with 4×4×8 patches and tube masking.

| Well | Well C | Well D | Well E | Well F | Well G | Well H | Well I | Well J |
|---|---|---|---|---|---|---|---|---|
| NMAE | $2.91\times10^{-2}$ | $5.42\times10^{-2}$ | $4.75\times10^{-2}$ | $2.91\times10^{-2}$ | $2.68\times10^{-2}$ | $2.72\times10^{-2}$ | $3.79\times10^{-2}$ | $2.72\times10^{-2}$ |
| NRMSE | $4.22\times10^{-2}$ | $7.04\times10^{-2}$ | $6.36\times10^{-2}$ | $4.31\times10^{-2}$ | $4.11\times10^{-2}$ | $3.84\times10^{-2}$ | $5.36\times10^{-2}$ | $3.76\times10^{-2}$ |
| PCC | 0.987 | 0.954 | 0.968 | 0.982 | 0.990 | 0.988 | 0.975 | 0.989 |
| Average NMAE: $3.49\times10^{-2}$; NRMSE: $4.87\times10^{-2}$; PCC: 0.979 | | | | | | | | |

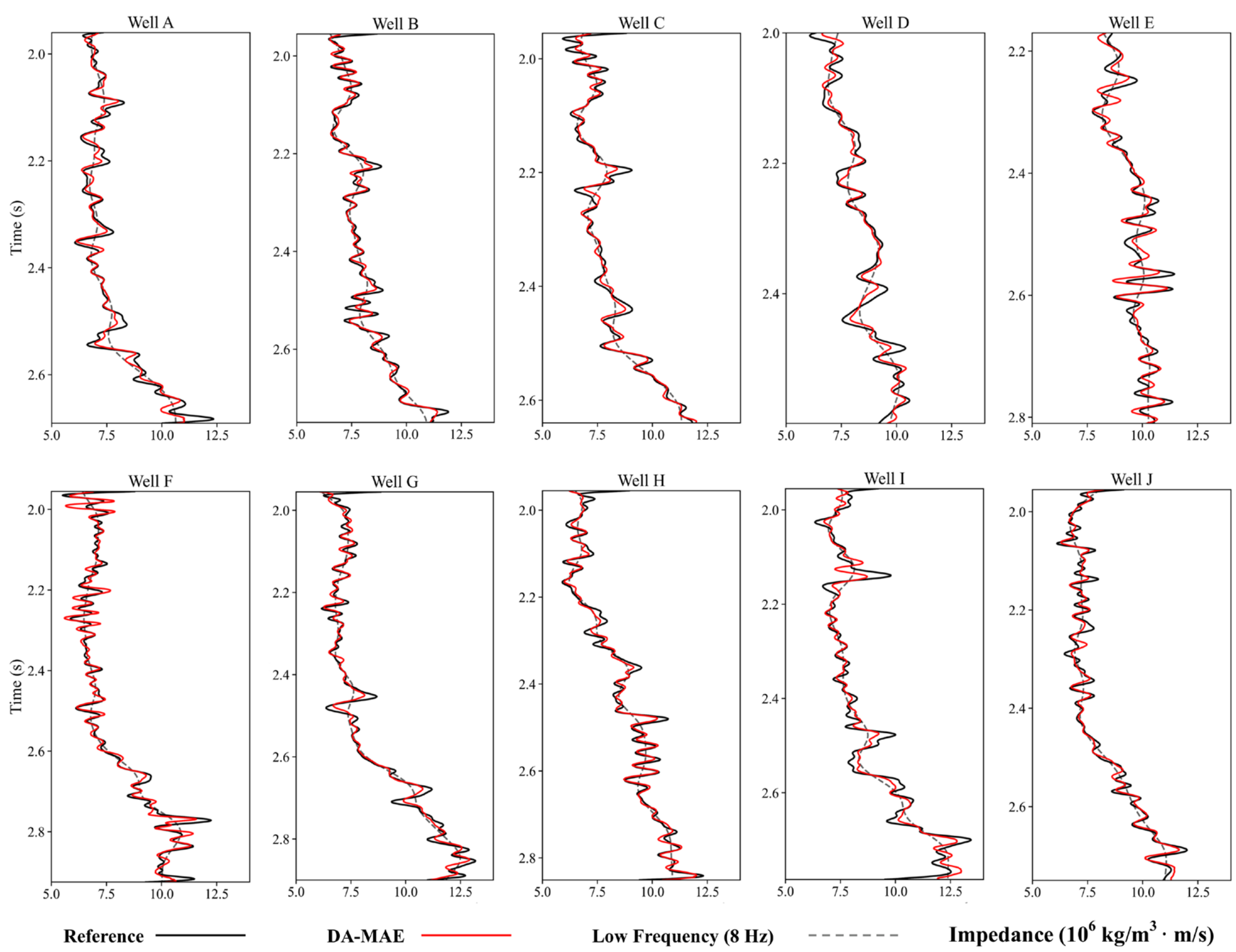

**Figure 8** Well-level acoustic impedance predictions in Area L obtained using the Full DA-MAE encoder with 4×4×8 patches and tube masking. The gray dashed, black, and red curves denote the 8-Hz background model, reference impedance, and predicted impedance, respectively.

### 5.2. Interwell impedance section prediction

To examine whether the well-level improvements extend to spatially coherent interwell predictions, the predicted center traces were assembled along a representative profile in Area L. Fig. 9 compares the input seismic section, the 8-Hz low-frequency impedance background, and the impedance predictions from different pretrained models.

All pretrained configurations recover impedance variations beyond the smooth low-frequency background, with the main lateral changes broadly following the seismic reflection geometry. The predicted impedance also remains consistent with the available well logs along the profile. These results indicate that the seismic representations learned during masked pretraining can be transferred to downstream impedance inversion and retain information useful for describing stratigraphic variation and lateral changes in subsurface properties. The 4×4×8 configurations preserve finer lateral and vertical impedance variations, whereas the 8×8×30 predictions are generally smoother. This is consistent with the well-level results and further suggests that token scale affects the amount of local seismic information retained for inversion. However, the visual differences among the pretrained models remain modest at the section scale. Therefore, the well-level quantitative results provide the primary evidence for differences in transfer performance.

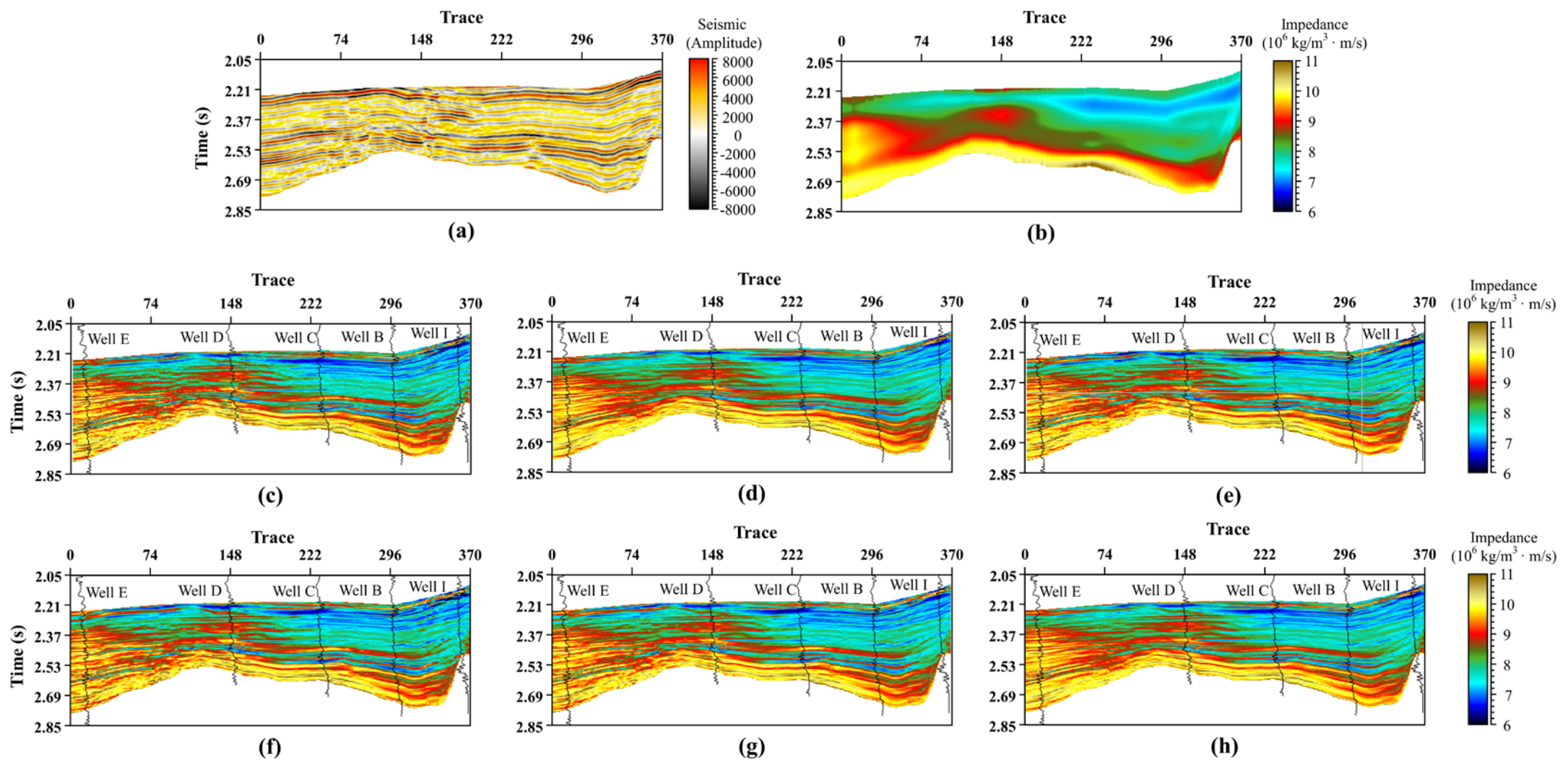

**Figure 9** Interwell acoustic impedance predictions in Area L. Panels (a) and (b) show the input seismic section and the 8-Hz background model, respectively. Panels (c)–(e) show predictions from the Backbone, tube-masked DA-MAE (Full), and random-masked DA-MAE models with 4×4×8 patches, while panels (f)–(h) show the corresponding predictions with 8×8×30 patches. Reference impedance logs are overlaid at the available well locations in panels (c)–(h).

## 6. Discussion

The results show that the effectiveness of masked seismic pretraining depends not only on reconstruction accuracy, but also on how token scale, masking strategy, and representation design match the requirements of the downstream inversion task. The following discussion focuses on the directional scale of seismic representation, the relationship between reconstruction and inversion, and the practical scope of the proposed framework.

### 6.1. Directional Token Scale and Seismic Context

The reconstruction experiments show a clear directional dependence on token scale. Increasing the lateral patch size causes each token to represent a broader spatial region, reducing the resolution available for local reflector geometry and trace-to-trace variations. Reconstruction is therefore more sensitive to lateral token resolution than to moderate changes in vertical patch length. In contrast, moderate vertical aggregation can retain useful waveform context while substantially reducing the number of tokens.

These results support anisotropic tokenization, in which lateral structural resolution and vertical waveform context are controlled separately.

The scaling experiments further show that increasing encoder capacity does not recover the information lost through coarse tokenization. This suggests that local seismic structure removed at the tokenization stage cannot simply be compensated for by a larger network. Token geometry should therefore be selected before model capacity is increased. The preferred vertical context also depends on the downstream task: some longer vertical patches produce better impedance inversion despite lower reconstruction accuracy. One possible interpretation is that impedance inversion benefits from a broader vertical seismic context, whereas masked reconstruction places greater emphasis on recovering local waveform detail.

### 6.2. Reconstruction fidelity and inversion transferability

Reconstruction fidelity and downstream transferability are not consistently aligned. Random masking gives substantially better reconstruction than tube masking, but under fine tokenization, tube masking produces better cross-area impedance inversion. Random masking leaves nearby observations available and can therefore favor local interpolation. Trace-aligned tube masking removes the complete vertical token sequence at selected lateral locations, forcing the encoder to make greater use of neighbouring traces and laterally coherent reflection structure. This broader cross-trace information appears to be more useful for impedance inversion than the short-range context that favours masked reconstruction.

The cross-area results also show that the pretrained representations retain information beyond the smooth low-frequency impedance background. At the test wells, the predicted impedance follows both the regional trend and shorter-scale variations in the reference logs, while the interwell predictions broadly follow the seismic reflection geometry. Such impedance variations are important for delineating stratigraphic boundaries and lateral changes in subsurface properties and can provide a quantitative basis for subsequent reservoir and lithologic interpretation. Thus, reconstruction metrics should primarily be regarded as measures of pretext-task performance rather than direct measures of the geophysical value of the learned representation.

### 6.3. Practical implications and limitations

The present study focuses on 3D post-stack seismic data, for which the lateral dimensions primarily describe reflector structure and the vertical direction records seismic waveforms. The proposed directional design is therefore most directly applicable to post-stack representation learning. Extending the framework to prestack or higher-dimensional seismic data will require the representation to account for amplitude variations with incident angle and azimuth.

A second limitation concerns the physical meaning of token scale. The patch dimensions used in this study are defined in traces and samples, whereas spatial sampling, temporal sampling, and seismic bandwidth vary among surveys. The preferred patch sizes should therefore be interpreted as evidence of directional scale sensitivity rather than universal optimal values. A more physically consistent formulation could relate lateral token support to spatial sampling and seismic wavelength, and vertical token extent to temporal sampling and waveform bandwidth. Further evaluation across areas with larger geological and acquisition differences will also be necessary to establish the limits of cross-area transfer.

## 7. Conclusions

This study presents DA-MAE, a direction-aware masked pretraining framework for representation learning from 3D post-stack seismic data. The framework distinguishes lateral reflector structure from vertical waveform characteristics through its token representation, masking strategy, and reconstruction objective. Masked reconstruction and cross-area acoustic impedance inversion are used to evaluate both pretext-task performance and downstream transferability.

The experiments show that seismic reconstruction is more sensitive to lateral token resolution than to moderate changes in vertical extent, and that increasing encoder capacity does not fully compensate for coarse tokenization. More importantly, the token scales and masking strategies that favor reconstruction do not necessarily yield the best inversion performance. Cross-area inversion further shows that the pretrained representations retain seismic information useful for recovering impedance variations in a different field area.

These results suggest that seismic masked pretraining should not be evaluated by reconstruction accuracy alone. Effective representation learning requires token scale, masking strategy, and model design to be considered together with the characteristics of the seismic data and the requirements of the downstream inversion task. The proposed framework provides a practical basis for using large volumes of unlabelled seismic data to support acoustic impedance inversion and subsurface characterisation.

## Funding

This work was supported by the National Science and Technology Major Project of China for New Oil and Gas Exploration and Development (2024ZD1400102), the National Natural Science Foundation of China (U24B2020), the Key Technology for Geophysical Prediction of Ultra-Deep Carbonate Reservoirs (P24240), and the Shandong Province Postdoctoral Innovation Seed Fund (SDZZ-ZR-202501411).

## Conflict of interest statement

None declared.

## Data availability

The data that support the findings of this study are available from the corresponding author upon reasonable request.

## Appendix

### A.1. Backbone-level downstream transfer results

**Table A1** Backbone-level downstream transfer performance across pretraining objectives and token configurations.

| Encoder / initialization | Encoder architecture | Patch size | Pretraining objective | Masking | NMAE | NRMSE | PCC |
|---|---|---|---|---|---|---|---|
| Pretrained | Backbone | 8×8×30 | Amplitude-only | Tube | $4.47\times10^{-2}$ | $6.45\times10^{-2}$ | 0.965 |
| Pretrained | Backbone | 8×8×120 | Seismic | Tube | $4.03\times10^{-2}$ | $5.67\times10^{-2}$ | 0.973 |
| Pretrained | Backbone | 8×8×60 | Seismic | Tube | $4.12\times10^{-2}$ | $5.63\times10^{-2}$ | 0.973 |
| Pretrained | Backbone | 8×8×30 | Seismic | Tube | $4.60\times10^{-2}$ | $6.45\times10^{-2}$ | 0.966 |
| Pretrained | Backbone | 8×8×20 | Seismic- | Tube | $5.34\times10^{-2}$ | $7.58\times10^{-2}$ | 0.953 |
| Pretrained | Backbone | 8×8×10 | Seismic | Tube | $5.33\times10^{-2}$ | $7.50\times10^{-2}$ | 0.954 |
| Pretrained | Backbone | 8×8×8 | Seismic | Tube | $5.31\times10^{-2}$ | $7.47\times10^{-2}$ | 0.954 |
| Pretrained | Backbone | 4×4×8 | Seismic | Tube | $4.45\times10^{-2}$ | $6.15\times10^{-2}$ | 0.968 |

*Note:* All models use the same prediction head, low-frequency input, wavelet-conditioning procedure, inversion objective, and fine-tuning protocol.

### A.2. Downstream inversion objective

The downstream inversion network predicts a normalised log-impedance trace $\hat{z}$ from each 3D seismic window, with the low-frequency impedance model incorporated additively as a background component. Let $\mathbf{z}$ denote the corresponding reference normalised log-impedance trace. The training objective is

$$L_{\text{down}}=\lambda_{\text{LM}}L_{\text{LM}}+L_{\text{MB}}+\lambda_{\phi}L_{\phi}+\lambda_{\text{grad}}L_{\text{grad}}+\lambda_{\text{bp}}L_{\text{bp}}+\lambda_{\text{fwd}}L_{\text{fwd}}. \tag{A5}$$

Let $S_k\left(\cdot\right)$ denote a moving-average operator with kernel length $k$. The low-to-mid-scale term is

$$L_{\text{LM}}=\|S_9\left(\hat{\mathbf{z}}\right)-S_9\left(\mathbf{z}\right)\|_1 . \tag{A6}$$

Four moving-average kernels, $k\in\{5,9,13,21\}$, further define five complementary multiscale components

$$\begin{aligned}&B_1\left(z\right)=S_{21}\left(z\right), B_2\left(z\right)=S_{13}\left(z\right)-S_{21}\left(z\right), B_3\left(z\right)=S_9\left(z\right)-S_{13}\left(z\right),\\&B_4\left(z\right)=S_5\left(z\right)-S_9\left(z\right), B_5\left(z\right)=z-S_5\left(z\right),\end{aligned} \tag{A7}$$

and the corresponding multiband loss is

$$L_{\text{MB}}=\sum_{j=1}^{5}w_j\|B_j\left(\hat{\mathbf{z}}\right)-B_j\left(\mathbf{z}\right)\|_1 ,\ w=\left(0.05,\ 0.10,\ 0.40,\ 0.45,\ 0\right). \tag{A8}$$

Phase consistency is imposed on the third and fourth multiscale components using an amplitude-weighted cosine phase discrepancy,

$$L_{\phi}=L_{\phi}^{(3)}+0.8L_{\phi}^{(4)},\ L_{\phi}^{(j)}=\left\langle\omega_j\left(f\right)\left[1-\cos\Delta\phi_j\left(f\right)\right]\right\rangle_f , \tag{A9}$$

where $\Delta\phi_j\left(f\right)=\hat{\phi}_j\left(f\right)-\phi_j\left(f\right)$ and $\omega_j\left(f\right)\propto\left|Z_j\left(f\right)\right|^{1/2}$ is normalised by its mean over nonzero frequencies.

The multistep vertical-gradient term is

$$L_{\text{grad}}=\frac{1}{3}\sum_{s\in\{1,2,4\}}\|\Delta_s\hat{\mathbf{z}}-\Delta_s\mathbf{z}\|_1 ,\ \Delta_s z_i=z_{i+s}-z_i. \tag{A10}$$

A 40–80-Hz cosine-tapered FFT constraint with a 5-Hz transition width is additionally imposed as

$$L_{\mathrm{bp}}=\mathrm{MSE}\left[\mathrm{B}_{40-80}\left(\hat{\mathbf{z}}\right),\mathrm{B}_{40-80}\left(\mathbf{z}\right)\right]. \tag{A11}$$

For seismic forward consistency, the predicted normalised log-impedance is converted to a first-order reflectivity approximation, $\hat{r}_i=\frac{1}{2}\left(\hat{z}_i\text{-}\hat{z}_{i\text{-}1}\right)$, and convolved with the wavelet $\mathbf{w}$ to generate synthetic seismic $\hat{\mathbf{d}}=\mathbf{w}*\hat{\mathbf{r}}$. The forward-consistency term is

$$L_{\mathrm{fwd}}=\|\,\mathrm{N}\left(\hat{\mathbf{d}}\right)\text{-}\mathrm{N}\left(\mathbf{d}\right)\|_1+\left(1\text{-}\rho_{\max}\right), \tag{A12}$$

where $\mathrm{N}(\cdot)$ denotes trace-wise standardization and $\rho_{\max}$ is the maximum absolute cross-correlation over lags from −8 to 8 samples.

The weights used in all reported downstream experiments were

$$\lambda_{\mathrm{LM}}=5.0,\ \lambda_{\phi}=0.2,\ \lambda_{\mathrm{grad}}=0.1,\ \lambda_{\mathrm{bp}}=1.0,\ \lambda_{\mathrm{fwd}}=2.0. \tag{A13}$$

**A.3. Definitions of reconstruction metrics**

Reconstruction performance is evaluated using MSE, 3D SSIM, PCC, SF, PF, LGE, and TGE. Let $F=\mathrm{F}_3\left(X\right)$ and $\hat{F}=\mathrm{F}_3\left(\hat{X}\right)$, where $\mathrm{F}_3$ denotes the 3D Fourier transform and $F_k$ and $\hat{F}_k$ are the target and reconstructed spectral coefficients at frequency index $k$, respectively.

MSE is computed over all voxels in each reconstructed 3D block. PCC is calculated after vectorizing the reconstructed and target blocks, and the resulting values are averaged over all test blocks. The 3D SSIM is computed directly on the volumetric data using a 7×7×7 Gaussian window with a standard deviation of 1.5, and is likewise averaged over all test blocks.

SF is defined as the cosine similarity between the Fourier amplitude spectra:

$$\mathrm{SF}=\frac{\sum_k\left|\hat{F}_k\right|\left|F_k\right|}{\sqrt{\sum_k\left|\hat{F}_k\right|^2}\sqrt{\sum_k\left|F_k\right|^2}+\epsilon}. \tag{A1}$$

PF is calculated from the amplitude-weighted wrapped phase difference:

$$\mathrm{PF}=1-\frac{1}{\pi}\sum_k w_k\left|\Delta\phi_k\right| \tag{A2}$$

where $\Delta\phi_k=\arg\{\exp\left[i\left(\arg\hat{F}_k-\arg F_k\right)\right]\}$, $w_k=\frac{\left|\hat{F}_k\right|\left|F_k\right|}{\sum_j\left|\hat{F}_j\right|\left|F_j\right|+\epsilon}$. The amplitude-based weights reduce the contribution of frequency components with negligible energy. Both SF and PF range from 0 to 1, with higher values indicating better spectral and phase agreement.

To evaluate direction-dependent reconstruction errors, LGE measures the discrepancy between the inline and crossline first-order gradients:

$$\mathrm{LGE}=\frac{1}{2}\left[\left\langle\left|\Delta_H\hat{X}-\Delta_H X\right|\right\rangle+\left\langle\left|\Delta_W\hat{X}-\Delta_W X\right|\right\rangle\right], \tag{A3}$$

whereas TGE measures the corresponding discrepancy along the vertical time direction:

$$\mathrm{TGE}=\left\langle\left|\Delta_D\hat{X}-\Delta_D X\right|\right\rangle. \tag{A4}$$

Here, $\Delta_H$, $\Delta_W$, and $\Delta_D$ denote first-order forward-difference operators along the inline, crossline, and vertical axes, respectively, and $\langle\cdot\rangle$ denotes averaging over all valid elements. Lower LGE and TGE indicate better preservation of lateral reflector continuity and vertical waveform variations.

## References

Alfarraj M, AlRegib G. 2019. Semisupervised sequence modeling for elastic impedance inversion. Interpretation, 7: SE237–49. https://doi.org/10.1190/INT-2018-0250.1

Arnab A, Dehghani M, Heigold G, Sun C, Lučić M, Schmid C. 2021. ViViT: A video vision transformer. Proceedings of the IEEE/CVF International Conference on Computer Vision (ICCV): 6836–46. https://doi.org/10.1109/ICCV48922.2021.00676

Chopra S, Marfurt KJ. 2005. Seismic attributes—A historical perspective. Geophysics, 70: 3SO–28SO. https://doi.org/10.1190/1.2098670

Das V, Pollack A, Wollner U, Mukerji T. 2019. Convolutional neural network for seismic impedance inversion. Geophysics, 84: R869–80. https://doi.org/10.1190/geo2018-0838.1

Dou YM, Li KW. 2024. 3D seismic mask auto encoder: Seismic inversion using transformer-based reconstruction representation learning. Comput Geotech, 169: 106194. https://doi.org/10.1016/j.compgeo.2024.106194

He KM, Chen XL, Xie SN, Li YH, Dollár P, Girshick R. 2022. Masked autoencoders are scalable vision learners. Proceedings of the IEEE/CVF Conference on Computer Vision and Pattern Recognition (CVPR): 15979–88. https://doi.org/10.1109/CVPR52688.2022.01553

Khosro Anjom F, Vaccarino F, Socco LV. 2024. Machine learning for seismic exploration: Where are we and how far are we from the holy grail? Geophysics, 89: WA157–78. https://doi.org/10.1190/geo2023-0129.1

Latimer RB, Davidson R, van Riel P. 2000. An interpreter's guide to understanding and working with seismic-derived acoustic impedance data. Leading Edge, 19: 242–56. https://doi.org/10.1190/1.1438580

Li GL, Li YY, Huang JP, Wu XY. 2025. A pre-training and fine-tuning paradigm for building a subsurface model. J Geophys Eng, 22: 877–88. https://doi.org/10.1093/jge/gxaf044

Li YY, Alkhalifah T, Huang JP, Li ZC. 2023. Self-supervised pretraining Vision Transformer with masked autoencoders for building subsurface model. IEEE Trans Geosci Remote Sens, 61: 4506610. https://doi.org/10.1109/TGRS.2023.3308999

Liu SX, Birnie C, Bakulin A, Dawood A, Silvestrov I, Alkhalifah T. 2024. A self-supervised scheme for ground roll suppression. Geophys Prospect, 72: 2580–98. https://doi.org/10.1111/1365-2478.13522

Meng JY, Wang SD, Niu GH, Sang WJ, Geng WH, Cheng WL. 2024. Seismic impedance inversion using a multi-input neural network with a two-step training strategy. Geophys Prospect, 72: 107–24. https://doi.org/10.1111/1365-2478.13229

Oldenburg DW, Scheuer T, Levy S. 1983. Recovery of the acoustic impedance from reflection seismograms. Geophysics, 48: 1318–37. https://doi.org/10.1190/1.1441413

Sheng HL, Wu XM, Si X, Li JT, Zhang SB, Duan XD. 2025. Seismic foundation model: A next generation deep-learning model in geophysics. Geophysics, 90: IM59–79. https://doi.org/10.1190/geo2024-0262.1

Song L, Yin XY, Zong ZY, Jiang M. 2022. Semi-supervised learning seismic inversion based on spatio-temporal sequence residual modeling neural network. J Petrol Sci Eng, 208: 109549. https://doi.org/10.1016/j.petrol.2021.109549

Sun J, Innanen KA, Huang C. 2021. Physics-guided deep learning for seismic inversion with hybrid training and uncertainty analysis. Geophysics, 86: R303–17. https://doi.org/10.1190/geo2020-0312.1

Sun QH, Zong ZY. 2024. Building initial model for seismic inversion based on semi-supervised learning. Geophys Prospect, 72: 1800–15. https://doi.org/10.1111/1365-2478.13491

Sun QH, Zong ZY, Li X. 2024. Probabilistic seismic inversion based on physics-guided deep mixture density network. Pet Sci, 21: 1611–31. https://doi.org/10.1016/j.petsci.2023.12.015

Tong Z, Song YB, Wang J, Wang LM. 2022. VideoMAE: Masked autoencoders are data-efficient learners for self-supervised video pre-training. Adv Neural Inf Process Syst, 35: 10078–93. https://doi.org/10.52202/068431-0732

Tran D, Wang H, Torresani L, Ray J, LeCun Y, Paluri M. 2018. A closer look at spatiotemporal convolutions for action recognition. Proceedings of the IEEE/CVF Conference on Computer Vision and Pattern Recognition (CVPR): 6450–59. https://doi.org/10.1109/CVPR.2018.00675

Wu BY, Meng D, Wang LL, Liu NH, Wang Y. 2020. Seismic impedance inversion using fully convolutional residual network and transfer learning. IEEE Geosci Remote Sens Lett, 17: 2140–44. https://doi.org/10.1109/LGRS.2019.2963106

Wu K, Peng HW, Chen MH, Fu JL, Chao HY. 2021. Rethinking and improving relative position encoding for vision transformer. Proceedings of the IEEE/CVF International Conference on Computer Vision (ICCV): 10033–41. https://doi.org/10.1109/ICCV48922.2021.00988

Wu XY, Huang JP, Li GL, Zhang XT, Zhang SS, Tian K, Qin N, Duan WS. 2026. Seismic-well fusion velocity model building via masked autoencoders network: Application to a field 3D seismic data. J Appl Geophys, 248: 106137. https://doi.org/10.1016/j.jappgeo.2026.106137

Yilmaz Ö. 2001. Seismic Data Analysis: Processing, Inversion, and Interpretation of Seismic Data. Tulsa, OK: Society of Exploration Geophysicists. https://doi.org/10.1190/1.9781560801580

Zou BL, Wang YJ, Chen T, Liang JD, Yu G, Hu GM. 2024. The domain adversarial and spatial fusion semi-supervised seismic impedance inversion. IEEE Trans Geosci Remote Sens, 62: 5900615. https://doi.org/10.1109/TGRS.2023.3336392